\documentclass[12pt,titlepage,oneside]{revtex4}

\usepackage{psfrag,graphicx}

\usepackage{dcolumn}

\usepackage{amsmath,amssymb}

\usepackage{bm}

\usepackage[dvips]{color}

\definecolor{mygrey}{gray}{0.35}

\definecolor{mygreen}{rgb}{0.85,1,0.9}

\definecolor{myzard}{cmyk}{0,0,0.05,0}

\definecolor{mywhite}{rgb}{1,1,1}

\definecolor{myred}{rgb}{1,0,0}

 \def\ii{\mathord{\rm i}}

\def\vec#1{{\bf{#1}}}

 \def\ket#1{|#1\rangle}

\begin{document}

\title[Short Title]{
Dirac Cat States in Relativistic Landau Levels}

\author{A. Bermudez$^1$, M.A. Martin-Delgado$^1$ and E. Solano$^{2,3}$}

\affiliation{ $^1$Departamento de F\'{\i}sica Te\'orica I,
Universidad Complutense, 28040. Madrid, Spain \\$^2$ Physics
Department, ASC, and CeNS, Ludwig-Maximilians-Universit\"at,
Theresienstrasse 37, 80333 Munich, Germany \\$^3$Secci\'on
F\'isica, Departamento de Ciencias, Pontificia Universidad
Cat\'olica del Per\'u, Apartado Postal 1761, Lima, Peru }

\begin{abstract}

We show that a relativistic version of Schr\"{o}dinger cat states, here called {\it Dirac cat states},
can be built in relativistic Landau levels when an external magnetic field couples to a relativistic spin $1/2$
charged particle. Under suitable initial conditions,
the associated Dirac equation produces unitarily Dirac cat states involving the orbital
quanta of the particle in a well defined mesoscopic regime. We demonstrate that the proposed
 Dirac cat states have a purely relativistic origin and cease to exist in the non-relativistic limit.
 In this manner, we expect to open relativistic quantum mechanics to the rich structures of quantum optics and
 quantum information.

\end{abstract}


\maketitle

Schr\"{o}dinger cat states were introduced to highlight the
distinctive and fundamental properties of quantum mechanics as
opposed to classical theories \cite{schrodinger_cat}. However,
their reach has gone far beyond and, presently, different
subfields of quantum information, like quantum communication,
fault-tolerant quantum computation, secret sharing, among others,
use them as a fundamental resource \cite{NielsenChuang}. For a cat
state we shall understand a coherent superposition of two
maximally different quantum states with the additional property of
being mesoscopic. The construction of cat states based on
relativistic quantum effects has not been addressed so far and it
is one of the main purposes of our work. To achieve this goal, we
study the Dirac Hamiltonian under specific conditions that we
detail below. We want to stress that the multidisciplinary
description of novel quantum relativistic effects is not only an
important advance in physics fundamentals, it will also influence
other physical systems that simulate their dynamics.

A relativistic electron of mass $m$, charge $-e$, subjected to a
constant homogeneous magnetic field along the $z$-axis, is
described by means of the Dirac equation
\begin{equation}
\ii \hbar \frac{\partial \ket{\Psi}}{\partial
t}=\left(c\boldsymbol{\alpha}(\textbf{p}+e\textbf{A})+mc^2\beta\right)\ket{\Psi},
\label{dirac_equation}
\end{equation}
where $\ket{\Psi}$ stands for the Dirac 4-component spinor,
$\textbf{p}$ represents the momentum operator, and $c$ the speed
of light. Here, $\vec{A}$ is the vector potential related to the
magnetic field through $\vec{B}=\nabla\wedge\vec{A}$, and
$\beta=\text{diag}(\mathbb{I},-\mathbb{I}),\alpha_j=\text{off-diag}(\sigma_j,\sigma_j)$
are the Dirac matrices in the standard representation with
$\sigma_j$ as the usual Pauli matrices \cite{greiner_book}. The
energy spectrum of this system is described by the relativistic
Landau levels, first derived by Rabi~\cite{rabi}
\begin{equation}
\label{landau_levels} E = \pm \sqrt { m^2 c^4 + p_z^2 c^2 + 2mc^2
\hbar \omega_c (n+1)},
\end{equation}
where $n=0,1,...$ and $\omega_c=eB/m$ is the cyclotron frequency
which describes the electron helicoidal trajectory.

In this paper, we derive an exact mapping between this
relativistic model and a combination of Jaynes-Cummings~(JC) and
Anti-Jaynes-Cummings~(AJC) interactions~\cite{jaynes_cummings}, so
widely used by the Quantum Optics community. This original
perspective allows a deeper understanding of relativistic
effects~\cite{lippmann}, as well as the prediction of novel
effects such as the existence of Dirac cat states. These
paradigmatic states  constitute the relativistic extension of the
usual Schr\"{o}dinger cat states~\cite{schrodinger_cat}. In the
same spirit as the latter, the Dirac cats involve a coherent
superposition of mesoscopically distinct states, but have a purely
relativistic nature.

\begin{figure}[!hbp]

\centering

\includegraphics[width=6.0 cm]{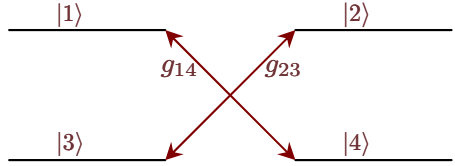}\\

\caption{Quantum Optical representation of the relativistic $e^-$
levels coupled by means of a constant magnetic
field.}\label{niveles}

\end{figure}

Working in the axial gauge, where
$\vec{A}:=\frac{B}{2}[-y,x,0]$,  the relativistic Hamiltonian can
be expressed as follows
\begin{equation}
\label{dirac_axial_gauge}
 H_{\text{D}}=mc^2\beta+\alpha_zp_z
+c\alpha_x(p_x-m\omega y)+c\alpha_y(p_y+m\omega x),
\end{equation}
where we have introduced $\omega:=\omega_c/2$. It is convenient to
introduce the chiral creation-annihilation operators
\begin{equation}
\label{circular_operators}
\begin{array}{c}
  a_r:=\frac{1}{\sqrt{2}}(a_x - \ii a_y),\hspace{2ex}a_r^\dagger:=\frac{1}{\sqrt{2}}(a_x^\dagger + \ii a_y^\dagger) , \\
  a_l:=\frac{1}{\sqrt{2}}(a_x + \ii a_y), \hspace{2ex}a_l^\dagger:=\frac{1}{\sqrt{2}}(a_x^\dagger - \ii a_y^\dagger) , \\
\end{array}
\end{equation}
where $ a_x^\dagger,a_x,  a_y^\dagger, a_y$, are the
creation-annihilation operators of the harmonic oscillator
$a^{\dagger}_i=\frac{1}{\sqrt{2}}\left(\frac{1}{\tilde{\Delta}}r_i
- \ii \frac{\tilde{\Delta}}{\hbar}p_i\right)$, $i=x,y$ and
$\tilde{\Delta}=\sqrt{\hbar/m\omega}$ represents the oscillator's
ground state width. Let us first consider an inertial frame
$\mathcal{S'}$ which moves along the axis $OZ$ at constant
$v_z=p_z/m$ with respect to a rest frame $\mathcal{S}$. In the
moving frame, the momentum becomes $p'_z=0$ in
Eq.~\eqref{dirac_axial_gauge}, and using these chiral
operators~\eqref{circular_operators}, the Dirac Hamiltonian
becomes
\begin{equation}
\label{dirac_hamiltonian_matrix}
H_{\text{D}}=mc^2\left[
\begin{array}{cccc}
  1                                 & 0                             & 0                                  & -\ii 2 \sqrt{ \xi } a_r \\
  0                                 & 1                             & \ii 2 \sqrt{ \xi } a_r^\dagger     & 0 \\
  0                                 &  -\ii 2 \sqrt{ \xi } a_r      & -1                                 & 0 \\
  \ii 2 \sqrt{ \xi } a_r ^\dagger   & 0                             & 0                                  & -1 \\
\end{array}
\right],
\end{equation}
where $\xi:=\hbar\omega/mc^2$ is a parameter which controls the
non-relativistic limit. It follows from
Eq.~\eqref{dirac_hamiltonian_matrix}, that the chiral operator
couples different components of the Dirac spinor and
simultaneously creates or annihilates right-handed quanta.
Expressing the Dirac spinor appropriately
 $|\Psi\rangle:=[\psi_1,\psi_2,\psi_3,\psi_4]^t$, the
 Hamiltonian becomes
\begin{equation}
\label{dirac_hamiltonian_JC+AJC}
\begin{split}
H_{\text{D}} = & mc^2 \sigma_{14}^z + g_{14}\sigma_{14}^+ a_r + g_{14}^*\sigma_{14}^- a_r^\dagger\\
+ & mc^2 \sigma_{23}^z + g_{23}\sigma_{23}^+a_r^\dagger +
g_{23}^*\sigma_{23}^-a_r ,
\end{split}
\end{equation}
where $g_{14}:=-\ii 2mc^2\sqrt{\xi}=:-g_{23}$ represent the
coupling constants between the different spinor components. The
first term in Eq.~\eqref{dirac_hamiltonian_JC+AJC} which couples
components $\{\psi_1,\psi_4\}$ is identical to a detuned
Jaynes-Cummings interaction
\begin{equation}
\label{dirac_hamiltonian_JC}
 H_{\text{JC}}^{14}=\Delta
\sigma_{14}^z + \left(g_{14}\sigma_{14}^+ a_r +
g_{14}^*\sigma_{14}^- a_r^\dagger\right).
\end{equation}
Likewise, the remaining term is identical to a
anti-Jaynes-Cummings (AJC) interaction between $\{\psi_2,\psi_3\}$
\begin{equation}
\label{dirac_hamiltonian_AJC}
H_{\text{AJC}}^{23}=\Delta
\sigma_{23}^z + \left(g_{23}\sigma_{23}^+ a_r^\dagger +
g_{23}^*\sigma_{23}^- a_r\right),
\end{equation}
with a similar detuning parameter $\Delta:=mc^2$. Therefore, the
Dirac Hamiltonian is the sum of JC and  AJC terms
$H_{\text{D}}=H_{\text{JC}}^{14}+H_{\text{AJC}}^{23}$, which is
 represented in  Fig.~\ref{niveles}.

This level diagram, so usual in Quantum Optics, must be
interpreted as follows. According to the free Dirac equation
$g_{14}=g_{23}=0$, the spinor components $\{\psi_1,\psi_2\}$
correspond to positive energy components, while
$\{\psi_3,\psi_4\}$ stand for negative energy components separated
by an energy gap $\Delta\epsilon=2mc^2$. Furthermore, these
components have  a well-defined value of the spin projected along
the $z-$axis. Namely, $\{\psi_1,\psi_3\}$ are spin-up components
while $\{\psi_2,\psi_4\}$ represent spin-down components. Thus, as
Fig.~\ref{niveles} states, the interaction of a free electron with
a constant magnetic field induces transitions between
spin-up/spin-down  and positive/negative energy components. Each
transition between the large and short components
$\{\psi_1,\psi_2\}\leftrightarrow \{\psi_3,\psi_4\}$ is
accompanied by a spin flip and mediated through the creation or
annihilation of right-handed quanta of rotation.

Taking advantage of usual methods in Quantum Optics, the whole
Hilbert space can be divided into a set of invariant subspaces,
which facilitate the diagonalization task. In order to do so, let
us introduce the states $|j,n_r\rangle=|j\rangle|n_r\rangle$,
which represent the electronic spinor component $\psi_j$ and the
electronic rotational state $
|n_r\rangle:=\frac{1}{\sqrt{n_r!}}(a_r^\dagger)^{n_r}|\text{vac}\rangle
$. Due to the previously described
mapping~\eqref{dirac_hamiltonian_JC+AJC}, the Hilbert space can be
described as
$\mathcal{H}=\mathcal{\tilde{H}}\bigoplus_{n_r=0}^{\infty}\mathcal{H}_{n_r}$,
where $\mathcal{\tilde{H}}$ is spanned by states
\begin{equation}
\mathcal{\tilde{H}}=\text{span}\{\ket{4,0},\ket{2,0}\},
\end{equation}
which have energies $ \tilde{E}:=\pm\Delta=\pm mc^2 $
respectively. These states can be interpreted as quantum optical
dark states, since they
 do not evolve exchanging chiral quanta~\eqref{dirac_hamiltonian_JC+AJC}. The remaining invariant
subspaces are
\begin{equation}
\label{subspace}
\mathcal{H}_{n_r}=\text{span}\{\ket{1,n_r},\ket{4,n_r+1},\ket{2,n_r+1},\ket{3,n_r}\}.
\end{equation}
and allow a block decomposition of the
Hamiltonian~\eqref{dirac_hamiltonian_matrix}
\begin{equation}
H_{n_r}=\left[
\begin{array}{cccc}
  \Delta          & -g\sqrt{n_r+1} & 0               & 0 \\
  -g^*\sqrt{n_r+1}  & -\Delta          & 0               & 0 \\
  0               & 0                &\Delta           & g\sqrt{n_r+1} \\
  0               & 0                & g^*\sqrt{n_r+1}  & -\Delta
  \\
\end{array}
\right],
\end{equation}
where $g=\ii 2 mc^2 \sqrt{\xi}$ is related to the coupling
constants introduced in Eq.~\eqref{dirac_hamiltonian_JC+AJC}. This
Hamiltonian can be block-diagonalized, yielding the following
energies
\begin{equation}
\label{energy_levels} E'=\pm
E'_{n_r}:=\pm\sqrt{\Delta^2+|g|^2(n_r+1)},
\end{equation}
which correspond to the relativistic Landau levels in
Eq.~\eqref{landau_levels} with $p_z=0$. In the non-relativistic
limit, where $E'_{n_r}=mc^2+\epsilon'_{n_r}$ such that
$\epsilon'_{n_r}\ll mc^2$, we find that the energy spectrum in
Eq.~\eqref{energy_levels} can be expressed as $
\epsilon'_{n_r}\approx\hbar\omega_c(n_r+1)$, which are the usual
Landau levels~\cite{landau}. The associated relativistic
eigenstates are
\begin{equation}
\label{landau_eigenstates}
\begin{split}
\ket{\pm E'_{n_r},1}:=&\alpha^{\pm}_{n_r}\ket{n_r}\chi_{1\uparrow}\mp\ii\alpha^{\mp}_{n_r}\ket{n_r+1}\chi_{2\downarrow},\\
\ket{\pm
E'_{n_r},2}:=&\alpha^{\pm}_{n_r}\ket{n_r+1}\chi_{1\downarrow}\mp\ii\alpha^{\mp}_{n_r}\ket{n_r}\chi_{2\uparrow},
\end{split}
\end{equation}
where we have introduced the usual Pauli spinors
\begin{equation}
\chi_{1\uparrow}:=\left[
\begin{array}{c}
  1 \\
  0 \\
  0 \\
  0 \\
\end{array}
\right],
\chi_{1\downarrow}:=\left[
\begin{array}{c}
  0 \\
  1 \\
  0 \\
  0 \\
\end{array}
\right],
\chi_{2\uparrow}:=\left[
\begin{array}{c}
  0 \\
  0 \\
  1 \\
  0 \\
\end{array}
\right],
\chi_{2\downarrow}:=\left[
\begin{array}{c}
  0 \\
  0 \\
  0 \\
  1 \\
\end{array}
\right],
\end{equation}
and $\alpha^{\pm}_{n_r}:=\sqrt{(E'_{n_r}\pm mc^2)/2E'_{n_r}}$. The
rotational and spinorial properties of the eigenstates in
Eq.~\eqref{landau_eigenstates} become unavoidably entangled in the
moving inertial frame $\mathcal{S'}$.

\begin{figure}[!hbp]

\centering

\includegraphics[width=6.0 cm]{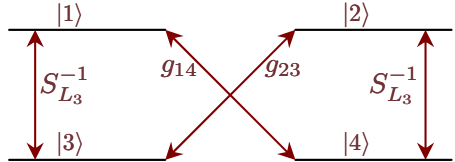}\\

\caption{Quantum Optical representation of the coupling between
the relativistic levels caused by the Lorentz
Boost.}\label{niveles_boost}

\end{figure}

 To obtain the
corresponding solutions in the rest frame $\mathcal{S}$, we must
perform a Lorentz boost along the $OZ$ axis $
p'^{\mu}:=[E'/c,p'^x,p'^y,0] \rightarrow
p^{\mu}:=[E/c,p^x,p^y,p^z]$. Considering the invariance of the
four-momentum
$g_{\mu\nu}p^{\mu}p^{\nu}=g_{\mu\nu}p'^{\mu}p'^{\nu}$, where the
Minkowski metric tensor is $g_{\mu\nu}=\text{diag}(1,-1,-1,-1)$ ,
and that $p^x=p'^x$,$p^y=p'^y$, we come to $
E'^2/c^2=E^2/c^2-p_z^2$. Substituting in Eq.~\eqref{energy_levels}
\begin{equation}
\label{energy_levels_rest}
 E=\pm E_{n_r}:=\pm\sqrt{\Delta^2+p_z^2c^2+|g|^2(n_r+1)}.
\end{equation}
These are the relativistic Landau levels in
Eq.~\eqref{landau_levels}, whose associated eigenstates may be
obtained by means of a Lorentz Boost to the Dirac spinor
$\Psi(x^{\mu})=S_{L_3}^{-1}\Psi'(x'^{\mu})$
\begin{equation}
\label{spinor_boost}
 S_{L_3}^{-1}=\text{cosh}\frac{\eta}{2}\left[
\begin{array}{cccc}
  1 & 0 & \text{tanh}\frac{\eta}{2}  & 0 \\
  0 & 1  & 0 & -\text{tanh}\frac{\eta}{2} \\
  \text{tanh}\frac{\eta}{2} & 0 & 1  & 0 \\
  0 & -\text{tanh}\frac{\eta}{2} & 0 & 1  \\
\end{array}
\right],
\end{equation}
where $\eta$ is the rapidity,
$\text{cosh}~\eta/2=\sqrt{(E_{n_r}+E'_{n_r})/2E'_{n_r}}$, $
\text{tanh}~\eta/2=p_zc/(E_{n_r}-E'_{n_r})$. With these
expressions, one can finally obtain the eigenstates in the rest
frame $\mathcal{S}$
\begin{equation}
\label{landau_eigenstates_rest}
\begin{split}
\ket{\pm E_{n_r},1}:=
 &\alpha^{\pm}_{n_r}\ket{n_r}\left(\text{cosh}\frac{\eta}{2}\chi_{1\uparrow}+\text{sinh}\frac{\eta}{2}\chi_{2\uparrow}\right)+ \\
 \pm&\ii\alpha^{\mp}_{n_r}\ket{n_r+1}\left(\text{sinh}\frac{\eta}{2}\chi_{1\downarrow}-\text{cosh}\frac{\eta}{2}\chi_{2\downarrow}\right),\\
 \ket{\pm E_{n_r},2}:=
 &\alpha^{\pm}_{n_r}\ket{n_r+1}\left(\text{cosh}\frac{\eta}{2}\chi_{1\downarrow}-\text{sinh}\frac{\eta}{2}\chi_{2\downarrow}\right)+ \\
 \mp&\ii\alpha^{\mp}_{n_r}\ket{n_r}\left(\text{sinh}\frac{\eta}{2}\chi_{1\uparrow}+\text{cosh}\frac{\eta}{2}\chi_{2\uparrow}\right),
\end{split}
\end{equation}
where the four spinor components get mixed in the rest frame
 $\mathcal{S}$ due to the Lorentz Boost (see Fig.~\ref{niveles_boost}).

Once the relativistic eigenstates have been obtained in a Quantum
Optics framework, we can discuss a novel aspect of the
relativistic electron dynamics, the rise of Dirac cat states. We
define the notion of Dirac cat states as a coherent superposition
of two mesoscopically distinct relativistic states. Our main goal
now is to find the conditions which guarantee the existence of
such cat states. They will turn out to be non-trivial. The mapping
of the Dirac Hamiltonian \eqref{dirac_equation} onto Quantum
Optics Hamiltonians \eqref{dirac_hamiltonian_JC+AJC} is a key tool
for finding the correct regime.

For the sake of simplicity we restrict to the regime with $p_z=0$,
where the effective dynamics of an initial state
$\ket{\Psi(0)}=|z_r\rangle\chi_{1\uparrow}$, with
$|z_r\rangle:=e^{-\frac{1}{2}|z_r|^2}\sum_{n_r=0}^{\infty}\frac{z_r^{n_r}}{\sqrt{n_r!}}\ket{n_r}$
being a right-handed coherent state with $z_r\in\mathbb{C}$, can
be described solely by the JC-term~\eqref{dirac_hamiltonian_JC}.
Due to the invariance of Hilbert subspaces, a blockade of the AJC
term occurs (see Fig.~\ref{niveles_2D}), and three different
regimes appear :

\begin{figure}[!hbp]

\centering

\includegraphics[width=6.0 cm]{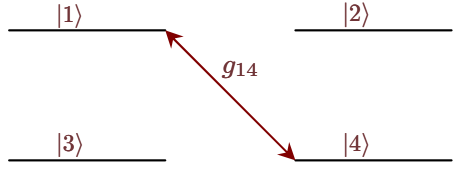}\\

\caption{Blockade of the AJC coupling}\label{niveles_2D}

\end{figure}

\vspace{1ex}

\textbf{Macroscopic Regime:}  In this regime, the mean number of
right-handed quanta $\bar{n}_r=|z_r|^2\to\infty$, so the
discreteness of the orbital degree of freedom can be neglected.
Setting $z_r=\ii|z_r|$, the JC-term~\eqref{dirac_hamiltonian_JC}
can be approximately described by the semiclassical Hamiltonian
\begin{equation}
H_{14}^{\text{sc}}=\Delta\sigma_z+|g||z_r|(\sigma^++\sigma^-),
\end{equation}
whose energies are $
 E^{\text{sc}}=\pm
E_{z_r}:=\pm\sqrt{\Delta^2+|g|^2|z_r|^2}. $ This semiclassical
energy levels  resemble the original
spectrum~\eqref{energy_levels}, but the corresponding eigenvalues
\begin{equation}
\label{eigenstates_semiclass}
 \ket{\pm E_{z_r}}:=\alpha^{\pm}_{z_r}\chi_{1\uparrow}\pm\ii\alpha^{\mp}_{z_r}\chi_{2\downarrow},
\end{equation}
with $\alpha^{\pm}_{z_r}:=\sqrt{(E_{z_r}\pm\Delta)/2E_{z_r}}$, are
clearly different from those in Eq.\eqref{landau_eigenstates}.
 In the semiclassical limit, entanglement between the spin and orbital
degrees of freedom is absent. The state
$\ket{\Psi(0)}:=\chi_{1\uparrow}$ evolves according to
\begin{equation}
\label{rabi_oscil_semiclass}
\begin{split}
|\Psi(t)\rangle = &
\left(\cos\Omega^{\text{sc}}_{z_r}t-\frac{\ii}{\sqrt{1+4\xi\bar{n}_r}}\sin\Omega^{\text{sc}}_{z_r}t\right)\chi_{1\uparrow}+ \\
& +
\ii\left(\sqrt{\frac{4\xi\bar{n}_r}{1+4\xi\bar{n}_r}}\sin\Omega^{\text{sc}}_{z_r}t\right)
 \chi_{2\downarrow},
\end{split}
\end{equation}
where $\Omega^{\text{sc}}_{z_r}:=E_{z_r}/\hbar$ is the
semiclassical Rabi frequency. Therefore, Dirac cats states of the
orbital degree of freedom cannot be produced during the dynamical
evolution.

\vspace{1ex}

\textbf{Microscopic Regime:} In this limit,
$\bar{n}_r=|z_r|^2\lesssim 10$ is small enough for the
discreteness of the orbital degree of freedom to become
noticeable. Especially interesting is the evolution of the vacuum
of right-handed quanta
\begin{equation}
\label{state_dynamics}
\begin{split}
|\Psi(t)\rangle = & \left(\cos\omega_{0}t-\frac{\ii}{\sqrt{1+4\xi}}\sin\omega_{0}t\right)|0\rangle \chi_{1\uparrow}+ \\
& + \left(\sqrt{\frac{4\xi }{1+4\xi}}\sin\omega_{0}t\right)
|1\rangle \chi_{2\downarrow},
\end{split}
\end{equation}
where $\omega_0:=  \frac{mc^2}{\hbar}\sqrt{1+4\xi}$ is the vacuum
Rabi frequency. We observe how the  spinorial and orbital degrees
of freedom become inevitably entangled as time evolves due to the
interference of positive and negative energy solutions, i.e.
\emph{Zitterbewegung}~\cite{dirac_2D}. This behavior is crucial
for the generation of Schr\"{o}dinger cat states, although their
growth cannot occur under this regime since the orbital degree of
freedom are not of a mesoscopic nature.

\vspace{1ex}

 \textbf{Mesoscopic Regime:} When the mean number
of orbital quanta $10\lesssim\bar{n}_r\lesssim100$ attains a
mesoscopic value , certain collapses and revivals in the Rabi
oscillations~\eqref{state_dynamics} occur~\cite{eberly}. An
asymptotic approximation
 which accounts for the collapse-revival phenomenon has been derived in~\cite{gea_banacloche_91,gea_banacloche_92},
 and its validity has been
experimentally tested in  Cavity QED (CQED)
~\cite{haroche_revival_cat}. Below, we derive a relativistic
 mesoscopic approximation, which allows us to
predict the generation of Dirac cat states.

Let us first discuss this asymptotic approximation, where the
semiclassical eigenstates~\eqref{eigenstates_semiclass} play an
essential role. The initial states $\ket{\Psi^{\pm}(0)}:=\ket{\pm
E_{z_r}}\ket{z_r}$ evolve according to
\begin{equation}
\label{asymptotic_approx} \ket{\Psi^{\pm}(t)}\approx
\left(\alpha^{\pm}_{z_r}e^{\mp\ii\frac{|g|^2}{2\hbar
E_{z_r}}t}\chi_{1\uparrow}\pm\ii\alpha^{\mp}_{z_r}\chi_{2\downarrow}\right)
e^{\mp\ii\Theta t}\ket{z_r},
\end{equation}
where
$\Theta:=\frac{1}{\hbar}\sqrt{\Delta^2+|g|^2a_r^{\dagger}a_r}$
depends on the chiral operators. The electron spin and orbital
degrees of freedom remain disentangled throughout the whole
evolution
$\ket{\Psi^{\pm}(t)}=\ket{\Phi_{\text{sp}}^{\pm}(t)}\otimes\ket{
\Phi_{\text{orb}}^{\pm}(t)}$. This peculiar behavior may be
compared to the  \emph{Zitterbewegung} oscillations  in
Eq.~\eqref{state_dynamics}, where entanglement plays a major role.

For times  shorter than the usual revival time $t\ll t_R:=2\pi
E_{z_r}\hbar/|g|^2$, the asymptotic approximation  in
Eq.\eqref{asymptotic_approx} can be pushed further, and a
suggestive expression for the evolved orbital state
$\ket{\Phi^{\pm}_{\text{orb}}(t)}:=e^{\mp\ii\Theta t}\ket{z_r}$
follows
\begin{equation}
\label{asymptotic_approx_orbital}
 \ket{\Phi_{\text{orb}}^{\pm}(t)}
 \approx
e^{\mp\ii\frac{t}{\hbar}\left(E_{z_r}-\frac{|g|^2|z_r|^2}{2E_{z_r}}\right)}|z_re^{\mp\ii\frac{|g|^2t}{2\hbar
E_{z_r}}}\rangle.
\end{equation}
Up to an irrelevant global phase, the short time evolution of the
orbital coherent state yields another coherent state whose phase
evolves in time according to
Eqs.~\eqref{asymptotic_approx_orbital}. Considering the position
operators $
X=\tilde{\Delta}(a_r+a_r^{\dagger}+a_l+a_l^{\dagger})/2$,
$Y=\ii\tilde{\Delta}(a_r-a_r^{\dagger}-a_l+a_l^{\dagger})/2$, we
calculate the expectation value that describes the electron
trajectory $\langle \textbf{X(t)} \rangle_{\pm}:=\left(\langle
X(t) \rangle_{\pm}\hspace{1ex},\langle Y(t) \rangle_{\pm}
\right)$, yielding the following

\begin{equation}
\label{circular_motion}
\begin{split}
\langle \textbf{X(t)} \rangle_+&=\tilde{\Delta}|z_r|\left(
-\sin\Omega_{\text{rot}}
t\hspace{1ex},+\cos\Omega_{\text{rot}} t\right),\\
\langle \textbf{X(t)} \rangle_-&=\tilde{\Delta}|z_r|\left(
+\sin\Omega_{\text{rot}} t\hspace{1ex},+\cos\Omega_{\text{rot}}
t\right),
\end{split}
\end{equation}
where $\Omega_{\text{rot}}:=|g|^2/2E_{z_r}\hbar$. Therefore
solutions $\ket{\Psi^{+}}$ rotate counterclockwise around the
$z-$axis, whilst $\ket{\Psi^{-}}$ rotate clockwise. Considering
$\ket{\Psi(0)}:=\chi_{1,\uparrow}|z_r\rangle=\left( \alpha^+_{z_r}
\ket{+E_{z_r}} + \alpha^-_{z_r} \ket{-E_{z_r}}\right)\ket{z_r}$,
which involves both semiclassical
solutions~\eqref{eigenstates_semiclass}, it splits up in two
components which rotate in opposite directions as time elapses
\begin{equation}
\label{state_dynamics_mesoscopic} \ket{\Psi(t)}= \alpha^+_{z_r}
\ket{\Phi_{\text{sp}}^+(t)}\ket{ \Phi_{\text{orb}}^+(t)} +
\alpha^-_{z_r} \ket{\Phi_{\text{sp}}^-(t)}\ket{
\Phi_{\text{orb}}^-(t)},
\end{equation}
where we have introduced the spinor states for clarity
\begin{equation}
\ket{\Phi_{\text{sp}}^{\pm}(t)}:=
\left(\alpha^+_{z_r}e^{\mp\ii\frac{|g|^2}{2\hbar
E_{z_r}}t}\chi_{1\uparrow}\pm\ii\alpha^-_{z_r}\chi_{2\downarrow}\right).
\end{equation}

Once we have discussed the relativistic asymptotic
approximation~\eqref{state_dynamics_mesoscopic}, we can proceed
with the generation of a relativistic version of Schr\"odinger cat
states. In order to obtain Dirac cats, we need the following
condition
\begin{equation}
\label{cat_condition}
\ket{\Phi_{\text{sp}}^+(t_d)}=e^{\ii\delta}\ket{\Phi_{\text{sp}}^-(t_d)}=:\ket{\tilde{\Phi}_d},
\end{equation}
to be fulfilled, where $t_d$ corresponds to the Dirac cat time and
$\delta\in\mathbb{R}$. If such a constraint~\eqref{cat_condition}
is satisfied, then the time
evolution~\eqref{state_dynamics_mesoscopic} under the mesoscopic
regime leads to
\begin{equation}
\label{schrodinger_cat} \ket{\Psi(t_d)}=\ket{\tilde{\Phi}_d}\left(
\alpha^+_{z_r} \ket{ \Phi_{\text{orb}}^+(t)} +
e^{i\delta}\alpha^-_{z_r} \ket{ \Phi_{\text{orb}}^-(t)}\right),
\end{equation}
and we obtain a  coherent superposition of states in the orbital
degree of freedom. Furthermore, using the properties of unitary
evolution, it follows that
\begin{equation}
\label{mesosc_disting}
\langle+E_{z_r}|-E_{z_r}\rangle=0\mapsto\langle\Phi_{\text{orb}}^+(t_d)|\Phi_{\text{orb}}^-(t_d)\rangle=0,
\end{equation}
and therefore the orbital state in Eq.\eqref{schrodinger_cat}
\begin{equation}
\ket{\Phi_{\text{orb}}^{\text{cat}}}:= \alpha^+_{z_r} \ket{
\Phi_{\text{orb}}^+(t_d)} + e^{i\delta}\alpha^-_{z_r} \ket{
\Phi_{\text{orb}}^-(t_d)},
\end{equation}
represents a coherent superposition of mesoscopically distinct
states, and consequently a Schr\"{o}dinger cat in the relativistic
scenario. The generation of these unusual cats is therefore
subjected to the verification of condition~\eqref{cat_condition}.
At half revival time $t_d=t_R/2=\pi E_{z_r}\hbar/|g|^2$, we find
\begin{equation}
\label{spinor_overlap}
|\langle\Phi_{\text{sp}}^+(t_d)|\Phi_{\text{sp}}^-(t_d)\rangle|\approx\sqrt{\frac{4\xi\bar{n}_r}{1+4\xi\bar{n}_r}}.
\end{equation}
In order to satisfy the aforementioned constraint, one must take
the ultra-relativistic limit $\xi\gg1/\bar{n}_r$, where
Eq.~\eqref{spinor_overlap} is
$|\langle\Phi_{\text{sp}}^+(t_d)|\Phi_{\text{sp}}^-(t_d)\rangle|\approx
1+\mathcal{O}(\frac{1}{\bar{n}_r})$ of the order of unity, and
thus a Dirac cat is generated. As a concluding remark, we stress
the relativistic nature of these cat states. In the
non-relativistic scenario Eq.\eqref{spinor_overlap} yields
\begin{equation}
|\langle\Phi_{\text{sp}}^+(t_s)|\Phi_{\text{sp}}^-(t_s)\rangle|\approx2\sqrt{\xi\bar{n}_r}+\mathcal{O}(\xi^{3/2})\ll1,
\end{equation}
and thus the cat generation condition cannot be fulfilled in this
case. As the electron slows down, the coherence of
~\eqref{schrodinger_cat} vanishes and the Dirac cat disappears.

In summary, we have found a novel correspondence between Quantum
Optics and Relativistic Quantum Mechanics. This perspective allows
an insightful derivation of the relativistic Landau levels, and
reveals a wide variety of original phenomena present in the
relativistic system. Remarkably, we have predicted the existence
of Dirac cat sates, a relativistic version of the unusual
Schr\"{o}dinger cat states, which have a purely relativistic
nature and occur under a mesoscopic regime.

\noindent {\em Acknowledgements} We acknowledge financial support
from a FPU M.E.C. grant (A.B.), EU project INSTANS (M.A.MD.), DGS
grant  under contract BFM 2003-05316-C02-01 and CAM-UCM grant
under ref. 910758 (A.B., M.A.MD.), and from DFG SFB 631, EU
EuroSQIP projects, and the German Excellence Initiative via the
``Nanosystems Initiative Munich (NIM)''.

\end{document}